\documentclass[pdflatex,sn-mathphys-num]{sn-jnl}

\usepackage{amsmath,amsfonts}
\usepackage{amssymb}
\usepackage{array}
\usepackage[caption=false,font=normalsize,labelfont=sf,textfont=sf]{subfig}
\usepackage{textcomp}
\usepackage{url}
\usepackage{verbatim}
\usepackage{natbib}

\usepackage{caption}
\usepackage{booktabs}
\usepackage{graphicx}
\usepackage{multirow}
\usepackage{bm}
\usepackage{xspace}

\usepackage{amsmath,amssymb,amsfonts}
\usepackage{algorithm}
\usepackage{graphicx}
\usepackage{textcomp}
\usepackage{xcolor}

\usepackage{graphicx}%
\usepackage{multirow}%
\usepackage{amsmath,amssymb,amsfonts}%
\usepackage{amsthm}%
\usepackage{mathrsfs}%
\usepackage[title]{appendix}%
\usepackage{xcolor}%
\usepackage{textcomp}%
\usepackage{manyfoot}%
\usepackage{booktabs}%
\usepackage{algorithm}%
\usepackage{algorithmicx}%
\usepackage{algpseudocode}%
\usepackage{listings}%

\theoremstyle{thmstyleone}%
\theoremstyle{thmstyletwo}%

\theoremstyle{thmstylethree}%

\begin{document}

\title[Article Title]{LD4MRec: Simplifying and Powering Diffusion Model for Multimedia Recommendation}


\author[1]{\fnm{Jiarui} \sur{Zhu}}\email{b21050320@njupt.edu.cn}

\author[2]{\fnm{Jun} \sur{Hou}}\email{houj.js@chinatelecom.cn}

\author[1]{\fnm{Penghang} \sur{Yu}}\email{2022010201@njupt.edu.cn}

\author*[1]{\fnm{Zhiyi} \sur{Tan}}\email{tzy@njupt.edu.cn}

\author[1]{\fnm{Bing-Kun} \sur{Bao}}\email{bingkunbao@njupt.edu.cn}

\affil*[1]{\orgname{Nanjing University of Posts and Telecommunications}, \orgaddress{\city{Nanjing}, \postcode{210003}, \state{Jiangsu}, \country{China}}}

\affil[2]{\orgname{China Telecom Jiangsu Branch}, \orgaddress{\city{Nanjing}, \postcode{210018}, \state{Jiangsu}, \country{China}}}


\abstract{Multimedia recommendation aims to predict users' future behaviors based on observed behaviors and item content information. However, the inherent noise contained in observed behaviors easily leads to suboptimal recommendation performance. Recently, the diffusion model's ability to generate information from noise presents a promising solution to this issue, prompting us to explore its application in multimedia recommendation. Nonetheless, several challenges must be addressed: 1) The diffusion model requires simplification to meet the efficiency requirements of real-time recommender systems, 2) The generated behaviors must align with user preference. To address these challenges, we propose a {\bfseries L}ight {\bfseries D}iffusion model for {\bfseries M}ultimedia {\bfseries Rec}ommendation ({\bfseries LD4MRec}). LD4MRec largely reduces computational complexity by employing a forward-free inference strategy, which directly predicts future behaviors from observed noisy behaviors. Meanwhile, to ensure the alignment between generated behaviors and user preference, we propose a novel {\bfseries C}onditional neural {\bfseries Net}work ({\bfseries C-Net}). C-Net achieves guided generation by leveraging two key signals, collaborative signals and personalized modality preference signals, thereby improving the semantic consistency between generated behaviors and user preference. Experiments conducted on three real-world datasets demonstrate the effectiveness of LD4MRec.}

\keywords{multimedia recommendation, diffusion model, conditional generation}



\maketitle

\section{Introduction}
Recommender systems have attained wide adoption in diverse domains. It make recommendations by analyzing observed user historical behaviors. Recognizing that user behaviors are influenced by item content information, recent research \cite{zhou2023bootstrap, yu2023multi} has focused on how to incorporate item content information to enhance recommendation performance.

Early recommendation researches \cite{he2016vbpr} are dominated in matrix factorization-based methods. These methods obtain behavioral features through matrix factorization, and combine them with item content features to model user preferences. Since user behaviors can be naturally represented as a bipartite graph, this has led to the emergence of Graph Neural Network (GNN)-based methods \cite{wei2019mmgcn,yu2023multi}. GNN-based methods are capable of capturing high-order collaborative signals within behavioral information \cite{he2020lightgcn}, enabling more effective modeling of user preferences. However, historical behaviors often contains noise in the form of false positives and false negatives. For example, users may mistakenly click on items they dislike. Such noise can result in suboptimal preference modeling. To address this issue, several self-supervised learning (SSL)-based methods \cite{tao2022self,wei2023multi} have been proposed. They introduce pretext task that maximize the agreement between behavioral representations under random perturbations, aiming to enhance robustness to such noise. Nevertheless, the random perturbations they introduce disrupt the already sparse behavioral semantic information, which may mislead user preference modeling and result in suboptimal performance.

\begin{figure}[!t]
  \centering
  \includegraphics[width=\linewidth]{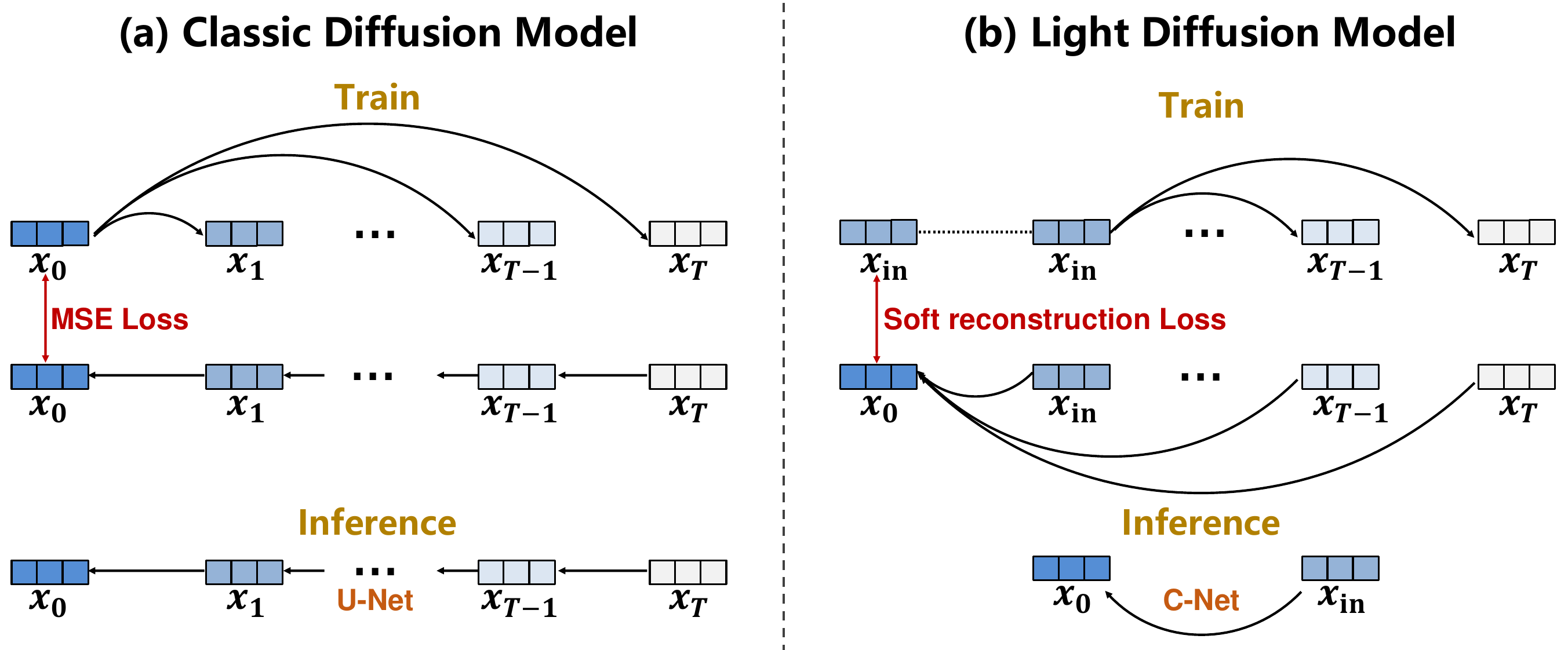}
  \caption{(a) Classic diffusion models generate data from Gaussian noise. (b) The proposed light diffusion model generates behaviors in a single step from observed noisy behaviors.}
  \label{fig.introduction}
\end{figure}

Recently, diffusion models have achieved remarkable success in Computer Vision (CV) and Natural Language Processing (NLP). These models typically consist of a forward process that corrupts input data by adding noise through a Markov chain, and a reverse process that learns to generate data from the corrupted data \cite{ho2020denoising}. Given that recommender systems aim to predict users' future behaviors based on observed noisy historical behaviors, the reverse process of diffusion models aligns well with the requirements. This motivates us to explore the use of diffusion models in multimedia recommendation. However, two critical challenges arise. First, recommender systems should handle large amount of data while ensuring timely responses \cite{wu2022survey}. Thus, it is essential to simplify the diffusion model to reduce computational complexity. Second, diffusion models often generate diverse information, that may not align with the requirements of the recommendation task. This necessitates the conditional generation to ensure that the generated behaviors meet user preference.

To address the aforementioned challenges, we propose a \textbf{L}ight \textbf{D}iffusion model for \textbf{M}ultimedia \textbf{Rec}ommendation (\textbf{LD4MRec}). On one hand, to reduce computational complexity, we design a forward-free inference strategy (as decipted in Fig. \ref{fig.introduction}). We experimentally demonstrate that it is unnecessary to corrupt historical behaviors during the inference phase, as recommendation tasks aim to predict future user behaviors rather than reproduce existing behaviors. Directly generating from existing behaviors not only preserves the valuable behavioral semantic information, achieving better recommendation results, but also largely reduce computational complexity. On the other hand, to ensure the generated behaviors meet user preference, we propose a novel \textbf{C}onditional neural \textbf{Net}work (\textbf{C-Net}) for the reverse process. C-Net employs a stacked architecture and incorporates two key guidance signals: collaborative signals and personalized modality preference signals. Collaborative signals reflect the co-occurrence pattern of behaviors, while the personalized modality preference signals reflect the causal pattern of behaviors. Generating future behaviors with the guidance of these two signals facilitates the generated behaviors aligning with user preferences. Additionally, since completely clean behavioral data is inaccessible, we utilize semi-supervised learning during C-Net training. By introducing a soft behavioral reconstruction constraint, C-Net is able to distill invariant user preferences and generate user future behaviors. Extensive experiments conducted on three real-world datasets demonstrate the effectiveness of LD4MRec. Our main contributions can be summarized: 

\textbf{$\bullet$} We propose a diffusion-based recommendation method, which mitigates the impact of noisy behaviors. 

\textbf{$\bullet$} We develop a novel C-Net for user future behaviors generation, which achieves effective generation under the guidance of collaborative and personalized modality preference signals.

\textbf{$\bullet$} We conduct extensive experiments on three real-world datasets, demonstrating the superiority of LD4MRec.

\section{Related Work}
\subsection{Multimedia Recommendation}
Multimedia recommender systems typically model user preferences by integrating content features with behavioral features. For example, MMGCN \cite{wei2019mmgcn} constructs multiple user-item graphs to mine behavioral features and combines these behavioral features with pre-extracted content features. Building on this foundation, LATTICE \cite{zhang2021mining} and MGCN \cite{yu2023multi} construct additional item-item graphs to exploit latent relationships among content features, thereby enhancing user preference modeling. However, existing methods often rely heavily on historical behavioral information, which can be problematic due to the presence of noise. Recently, although several self-supervised learning (SSL) methods, such as SLMRec \cite{tao2022self} and MMSSL \cite{wei2023multi}, have been proposed to mitigate this issue, the fundamental flaw remains unresolved.

\subsection{Diffusion models}
Diffusion models have demonstrated success in the generation tasks \cite{wallace2024diffusion}, notably in image synthesis. This success has prompted a number of studies to explore the application of diffusion models to various domains. However, there have been relatively few attempts to extend diffusion models to recommendation systems. The pioneer in designing diffusion models for recommendation purposes is CODIGEM \cite{walker2022recommendation}, which employs a set of distinct Autoencoders in an iterative manner for unconditional generation at each step. Building on this foundation, DiffRec \cite{wang2023diffusion} designs a shared Multilayer Perceptron (MLP) to achieve stable behavioral information generation, and DiffMM \cite{jiang2024diffmm} design a cross-modal contrastive learning paradigm to improve the generation. However, these methods fail to meet the efficiency requirements of recommender systems. Moreover, they ignore that the informational value density of behavioral information is extremely limited. The unconditional generation they used is hard to generate future behaviors that align with user preference. Thereby, they achieve suboptimal recommendation performance.

\section{Preliminary}
Let {$\mathcal{U}$} denote the set of users and {$\mathcal{I}$} denote the set of items. User historical behaviors is represented by {$\mathbf{R}$}, a binary matrix of size {$|\mathcal{U}| \times |\mathcal{I}|$}. Specifically, if user {$u$} has interacted with item {$i$} (e.g., click, purchase), $\mathbf{R}_{u,i} = 1$; otherwise, $\mathbf{R}_{u,i} = 0$. The objective of multimedia recommendation is to accurately predict the future interaction probabilities {$\hat{y}_{ui}$} between users and items, based on the observed historical behaviors $\mathbf{R}$.

\section{Light Diffusion Model}
\subsection{Forward and Reverse Processes}
Like most diffusion models, the proposed light diffusion model consists of two crucial processes. The first is the forward process, which gradually corrupts the historical behaviors through a Markov chain. The second is the reverse process, which learns to generate behaviors from corrupted behaviors.

\vspace{3pt}
\noindent\textbf{$\bullet$ Forward process.}
During forward process, diffusion model gradually corrupt the behaviors, simulating the user's mistakenly clicking behaviors. Considering the inherent noise already present in behavioral data, we do not assume that the input data is the ideal state data $\bm{x}_0$\footnote{For brevity, we omit the subscript $u$ in $\bm{x}_0$ and $\bm{x}_{\text{in}}$ for user $u$.}, as is common in other generation tasks. Instead, we assume that the input data has undergone a minor forward process, and we assign each user a parameter $t_u$ to indicate their individual forward diffusion step. Consequently, for user $u$, the behavioral data $\bm{r}_u$ can be represented as $\bm{x}_\text{in}$, which has undergone a forward process at step $t_u$. (To ensure that $t_u$ is learnable, we transform it into the initial step embedding $\textbf{z}_{u,in}$, similar to ID embedding.)

Next, we gradually corrupt the input data by adding noises through a stepwise Markov chain process over $T$ steps. Each step transition in this process is parameterized as follows:
\begin{equation}\label{eq:q_xt_given_xt-1}
    q(\bm{x}_t|\bm{x}_{t-1}) = \mathcal{N}(\bm{x}_t; \sqrt{1-\beta_t}\bm{x}_{t-1}, \beta_t\bm{I}),
\end{equation}
where $\mathcal{N}(x;\mu,\sigma^2)$ is a Gaussian distribution with a mean $\mu$ and variance $\sigma^2$, $\bm{x}_t$ is sampled from this Gaussian, $\beta_t$ is the noise added at the $t$-th diffusion step and $\bm{I}$ is the identity matrix. The value of $\beta_t$ is generated from a pre-defined noise schedule $\beta$ controlling the scale of Gaussian noise. 

By utilizing the reparameterization trick \cite{ho2020denoising} and the additivity property of two independent Gaussian noises \cite{ho2020denoising}, we can directly obtain {$\bm{x}_t$ from $\bm{x}_{\text{in}}$}:
\begin{equation}\label{eq:q_xt_given_xin}
\begin{aligned}
    q(\bm{x}_t|\bm{x}_{\text{in}}) &= \mathcal{N}(\bm{x}_t; \sqrt{\bar{\alpha}_t}\bm{x}_{\text{in}}, (1-\bar{\alpha}_t)\bm{I}) \\
    &= \sqrt{\bar{\alpha}_t} \bm{x}_{\text{in}} +  \sqrt{1-\bar{\alpha}_t}\epsilon,\thinspace\thinspace\thinspace \epsilon \sim \mathcal{N}(0, \bm{I}),
\end{aligned}
\end{equation}
where {$\alpha_t=1-\beta_t$}, $\bar{\alpha}_t = \prod_{i=1}^n \alpha_i$.

Since $\bar{\alpha}_t$ is a predefined noise schedule, we can effectively control $\bm{x}_t$ by establishing an appropriate $\beta_t$. In light of the information value density of user behavioral data is limited, we regulate the extent of noise introduced as proposed by \cite{wang2023diffusion}. To this end, a linear noise schedule is employed for $1-\bar{\alpha}_t$:
\begin{equation}
    1 - \bar{\alpha}_t = s\cdot\left[\alpha_{\min} + \frac{t-1}{T-1}(1 - \alpha_{\min})\right],\quad t\in{1,\dots,T}.
\end{equation}

Here, $s \in [0,1]$ serves as a hyperparameter governing the magnitude of the noise, while $\alpha_{\min} \in (0,1)$ is a hyperparameter indicating the minimum level of noise to be added. $T$ denotes the total number of forward diffusion steps. Normally, $s=0.1$ and $\alpha_{\min}=0.0001$ \cite{wang2023diffusion}.

\vspace{3pt}
\noindent\textbf{$\bullet$ Reverse process.}
During the reverse process, diffusion model learns to generate behaviors from the corrupted behaviors. This process enables the model to capture complex user-item interaction patterns while mitigating the negative impact of noise by modeling data uncertainty. For the corrupted data $x_t$, in addition to the noise added by the forward diffusion process, the input behaviors contains inherent noise. Therefore, the actual forward diffusion step is represented as $t' = t + t_u$.

Following \cite{jiang2024diffmm,wang2023diffusion}, we employ a neural network to predict the ideal state data $\bm{x}_0$ from the corrupted data $\bm{x}_t$. However, due to the low informational value density in behaviors, relying solely on corrupted behaviors presents challenges in effectively generating future behaviors that align with user preferences. To address this, we design a \textbf{C}onditional neural \textbf{Net}work (\textbf{C-Net}), which predicts user future behaviors $\hat{\bm{x}}_0$ with the guidance of additional conditioning information (abbreviated as "cond"). Formally, the C-Net is represented as:
\begin{equation}
\label{eq:diff_predict}
\begin{aligned}
    \hat{\bm{x}}_0 = p_{\theta}(\bm{x}_t,t',\text{cond}). \\
\end{aligned}
\end{equation}

\subsection{Efficient Training and Inference}
\vspace{3pt}
\noindent\textbf{$\bullet$ Training.}
\begin{algorithm}[t]
	\caption{\textbf{Efficient Training}}  
	\label{algo:training}
	\begin{algorithmic}[1]
		\Require All user historical behavioral data $\mathcal{R}$ (each user historical behaviors denote as $\bm{x}_{\text{in}}$), the condition (abbreviated as "cond") and randomly initialized parameters $\theta$.
            \Repeat 
            \For{a minibatch ${\mathbf{R}_n}\subset\mathbf{R}$}
            \State Sample the forward diffusion step $t\sim p_t$, 
            \State Sample random Gaussian noise $\bm{\epsilon}\sim\mathcal{N}(\bm{0},\bm{I})$;
            \State \textcolor{gray}{// \emph{Forward Process}}
            \State Compute $\bm{x}_t$ via $q(\bm{x}_t|\bm{x}_{\text{in}})$ in Eq. \ref{eq:q_xt_given_xin}
            \State \textcolor{gray}{// \emph{Reverse Process}}
            \State Calculate the actual foward diffusion step $t'$
            \State Predict $\bm{x}_0$ via $p_{\theta}(\bm{x}_t,t',\text{cond})$ in Eq. \ref{eq:diff_predict}
            \State \textcolor{gray}{// \emph{Optimization}}
            \State Calculate $\mathcal{L}_\text{rec}'$ via Eq. \ref{L_u}
            \State Take gradient descent step on $\nabla_\theta\mathcal{L}_\text{rec}'$
            \State \textcolor{gray}{// \emph{Dynamic Update}}
            \State Update the history value for $\mathcal{L}_\text{rec}'$ 
            \EndFor
            \Until{Converged}
        \Ensure Optimized parameters $\theta$.
	\end{algorithmic}
\end{algorithm}
In order to optimize the conditional neural network $\hat{\bm{x}}_{\theta}(\cdot)$, we enforce the generated behaviors $\hat{\bm{x}}_0$ to approximate the ideal behaviors $\bm{x}_0$:
\begin{equation}\label{eq:L_t_x}
    \begin{aligned}
        \mathcal{L} = \parallel p_{\theta}(\bm{x}_t,t',\text{cond})-\bm{x}_0\parallel_2^2.
    \end{aligned}
\end{equation}

However, due to inherent noise in the observed behaviors, the ideal behaviors $\bm{x}_0$ is not inaccessible. To address this, we propose a semi-supervised learning approach that incorporates soft labels. This approach aims to generate behavior that closely resembles observed historical behavioral data, while mitigating the impact of noisy behaviors and facilitates the generation of potential future behaviors. The modified loss function is represented as follows:
\begin{equation}
    \begin{aligned}
        \mathcal{L}_\text{rec} = \parallel p_{\theta}(\bm{x}_t,t',\text{cond})-f_s(\bm{x}_\text{in})\parallel_2^2.
    \end{aligned}
\end{equation}

Specifically, for the input observed behavioral data $\bm{x}_{\text{in}}$, we sample two subsets, $\mathcal{S}^+$ and $\mathcal{S}^-$, with a probability of $p$. The subset $\mathcal{S}^+$ represents the items that have been clicked on by the user and are thus labeled as 1, while the subset $\mathcal{S}^-$ represents the items that have not been clicked on and are labeled as 0. Then the smoothing operation is performed, as shown below:
\begin{equation}
    f_\text{{s}}(x)=\left\{
	\begin{aligned}
        &&1-\gamma,\quad&x \in \mathcal{S}^+,\\
        &&\gamma,\quad&x \in \mathcal{S}^-,\\
	\end{aligned}
	\right
        .
\end{equation}
where the hyperparameter $\gamma$ is employed to control the intensity of the smoothing effect.

When the forward diffusion step $t$ is relative small, the introduction of noise is minimal, allowing the model to effortlessly restore $\bm{x}_{0}$. Conversely, as $t$ approaches the total diffusion step $T$, the denoising of $\bm{x}_t$ becomes increasingly challenging due to its similarity to Gaussian noise. It is intuitive to allocate more training steps to more difficult samples, as this enhances the denoising capability. Therefore, we utilize the importance sampling \cite{nichol2021improved} to incentivize the inclusion of harder samples. The weighted loss function $\mathcal{L}_\text{rec}'$ is defined as:
\begin{equation}\label{L_u}
\small
    \mathcal{L}_\text{rec}'=\mathbb{E}_{n\sim n_t}\left[\frac{\mathcal{L}_{\text{rec}}}{n_t}\right], \quad n_t \propto \sqrt{\mathbb{E}[\mathcal{L}_{\text{rec}}]},\quad \sum_{t=1}^{T} n_t=1,
\end{equation}
where $n_t$ denotes the probability of sampling the forward diffusion step $t$. We dynamically record the value of each $\mathcal{L}_{\text{rec}}$ and update the record history accordingly. Before acquiring enough $\mathcal{L}_{\text{rec}}$, we adopt the uniform sampling. This training strategy is summarized in Algorithm \ref{algo:training}.

\vspace{3pt}
\noindent\textbf{$\bullet$ Inference.}
\begin{algorithm}[t]
	\caption{\textbf{Efficient Inference}}  
	\label{algo:inference}
	\begin{algorithmic}[1]
		\Require User historical behavioral data $\bm{x}_\text{in}$, the corresponding forward diffusion step $t_u$, the condition (abbreviated as "cond") and the optimized parameters $\theta$.
            \State \textcolor{gray}{// \emph{Reverse Process}}
            \State Predict $ \hat{\bm{x}}_0$ via the optimized network $p_{\theta}(\bm{x}_\text{in},t_u,\text{cond})$ 
        \Ensure Interaction probabilities $ \hat{\bm{x}}_0$ for user $u$.
	\end{algorithmic}
\end{algorithm}
The primary goal of recommendation is to predict users' future behaviors rather than generate what the user has already clicked. Thus, it is unnecessary to first corrupt the historical behaviors and then reconstruct the behaviors, as is commonly used in other generation tasks. Instead, we propose a novel forward-free inference strategy that directly predicts user future behaviors $\bm{x}_0$ from their historical behaviors $\bm{x}_\text{in}$. This not only reduces the computational costs, but also preserves the meaningful behavioral semantic information and achieves superior performance. In-depth experiments verify the rationality of our motivation. We summarize the forward-free inference strategy in Algorithm \ref{algo:inference}.

\subsection{Discussion}
To further investigate the effectiveness of the proposed forward free inference strategy, we conduct a theoretical analysis from the perspective of Bayesian theory. The performance of the diffusion model is evaluated in relation to Evidence Lower Bound (ELBO) \cite{Saharia2022Photorealistic}, a higher ELBO indicates improved model performance. The ELBO is calculated as follows:

\begin{equation}
\small
\text{ELBO} = \underbrace{\mathbb{E}_q[\log p(\bm{x}_t,H_u)]}_{\text{Prior Constraint}} - \underbrace{D_\text{KL}(q(\bm{x}_t|T_u,H_u) | p(\bm{x}_t,H_u))}_{\text{Initial Gap}} + \sum_{t=1}^T \underbrace{\mathbb{E}_q\left[\log\frac{p_\theta(\bm{x}_{t-1}|\bm{x}_t,H_u)}{q(\bm{x}_t|\bm{x}_{t-1},H_u)}\right]}_{\text{Step-wise Consistency}}
\end{equation}
where $\bm{x}_t$ is the data undergone $t$ forward process, $H_u$ is the observed interaction data of user $u$, $T_u$ represents the interaction data requiring prediction. Prior constraint measures alignment with the data prior, initial gap is the KL divergence between initial states, step-wise consistency ensures coherence between adjacent diffusion steps.

When introducing noise $\epsilon$ to $H_u$, the variational distribution becomes $q_\epsilon(\bm{x}_t|T_u,H_u) = \mathbb{E}_\epsilon [q(\bm{x}_t|T_u,H_u+\epsilon)]$. This causes entropy expansion:
\begin{equation}
H(q_\epsilon) \geq H(q) + H(\epsilon)
\end{equation}
The noisy distribution $q_\epsilon$ has increased entropy, widening its divergence from the target distribution.

Meanwhile, each step-wise term becomes:
\begin{equation}
\mathbb{E}{q_\epsilon}\left[\log\frac{p_\theta}{q_\epsilon}\right] = -D_\text{KL}(q_\epsilon | p_\theta) + C
\end{equation}
Where $C$ is a constant. The introduced noise strictly increases the KL divergence at each step:
\begin{equation}
D_\text{KL}(q_\epsilon | p_\theta) \geq D_\text{KL}(q | p_\theta)
\end{equation}

Let $\text{ELBO}_\text{foward-free}$ and $\text{ELBO}_\text{common}$ denote the bounds for the proposed forward-free and common strategies respectively:

\begin{equation}
\begin{aligned}
    \Delta\text{ELBO} &= \text{ELBO}_\text{foward-free} - \text{ELBO}_\text{common} \\
    &= \sum_{t=1}^T \left[D_\text{KL}(q_\epsilon^{(t)} | p_\theta^{(t)}) - D_\text{KL}(q^{(t)} | p_\theta^{(t)})\right] \geq 0
\end{aligned}
\end{equation}

This non-negative gap shows the cumulative degradation effect across all diffusion steps. As a result, the common inference strategy may yield suboptimal outcomes during inference, and forward-free inference strategy is more reasonable and demonstrate greater potential for application.

\section{C-Net}
\begin{figure*}
  \centering
  \includegraphics[width=\linewidth]{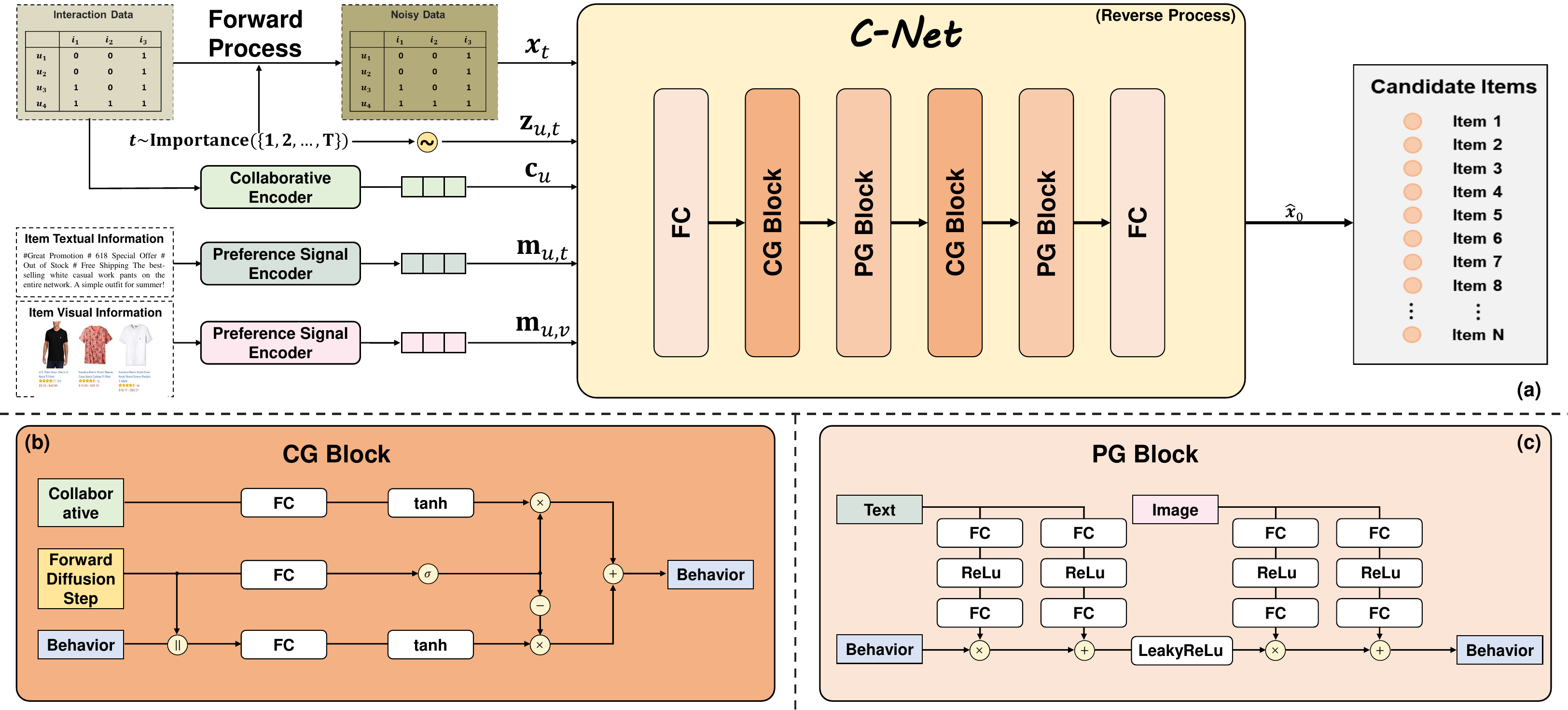}
  \caption{(a) The overall framework of LD4MRec. (b) CG-Block denoises the representations with the guidance of collaborative signals. (c) PG-Block generate behavior information under the control of user multimodal preferences.}
  \label{fig.The overall framework of LD4MRec}
\end{figure*}

$\hat{\bm{x}}_{\theta}(\cdot)$ in Eq. \ref{eq:diff_predict} represents the \textbf{C}onditional neural \textbf{Net}work, which generates user future behaviors that align with user preferences. This is illustrated in Figure \ref{fig.The overall framework of LD4MRec}. To maintain simplicity, we omit the depiction of nonlinear activation functions between the CG Blocks and PG Blocks.

\subsection{Dual Condition Signals}
To effectively generate user future behaviors, we adopt two key signal as the condition signals: collaborative signals and personalized modality preference signals. Collaborative signals leverage co-occurrence patterns among users to capture similar behavioral trends. In contrast, personalized modality preference signals aim to utilize the casual pattern of behaviors to simulate the user behaviors.

\vspace{3pt}
\noindent\textbf{$\bullet$ Collaborative Signals.}
We employ the randomized SVD \cite{rangarajan2001learning} scheme as the collaborative encoder to extract collaborative signals. Formally, it is represented as:
\begin{equation}
\begin{aligned}
\mathbf{U}, \bm{\Lambda}, \mathbf{I}^\intercal = \text{ApproxSVD}(\mathbf{R}, d_\text{svd}),\\
\end{aligned}
\end{equation}
where $\bm{\Lambda}$ is eigenvalues matrix, $\mathbf{U}$/$\mathbf{I}$ is the user/item eigenvector matrix, and $d_\text{svd}$ is the rank of decomposed matrix. To encode high-order co-occurrence patterns, we fuse the user eigenvector with clicked item eigenvector:
\begin{equation}
    \mathbf{C} = [\mathbf{U};\mathbf{D}_U^{-\frac{1}{2}} \mathbf{R} \mathbf{D}_I^{-\frac{1}{2}}\mathbf{I}],
\end{equation}
where $[\cdot;\cdot]$ is concatenation operation, $\mathbf{D}_U = \text{Diag}(\mathbf{R} \cdot \bm{1})$ and $\mathbf{D}_I = \text{Diag}(\bm{1}^\intercal \cdot \mathbf{R})$ are degree matrices, which avoids the scale of eigenvector increasing. The eigenvector $\textbf{c}_u$ is the $u$-th row of $\mathbf{C}$, which represents the collaborative signal of user $u$.

\vspace{3pt}
\noindent\textbf{$\bullet$ Personalized Modality Preference Signals.}
We employ a lightweight GCN module \cite{yu2023multi} to extract personalized modality preference signals. Taking the textual modality preference signal $\textbf{m}_{u,t}$ of user $u$ as an example:
\begin{equation}
\begin{aligned}
    \mathbf{m}_{u,t} = \sum_{i\in\mathcal{N}_u}\frac{1}{\sqrt{\lvert\mathcal{N}_u\rvert\lvert\mathcal{N}_i\rvert}}\mathbf{m}_{i,t},
\end{aligned}
\end{equation}
where $\mathbf{m}_{i,t}$ is the pre-extracted textual content features of item $i$, {$\mathcal{N}_u$} is the set of clicked items by the user ${u}$, and {$\mathcal{N}_i$} is the set of users who clicked item $i$. User visual modality preference signals $\textbf{m}_{u,v}$ can be obtained in the same way.

\subsection{Forward Diffusion Step Embeddings}
The forward diffusion step embedding is used to indicate the amount of noise added to the data, allowing the model to capture subtle variations in the generation process. We use the sinusoidal function \cite{Vaswani2017attention} to obtain the forward diffusion step embedding $\mathbf{z}_{u,t}\in\mathbb{R}^{d_\text{forward}}(1{\leq}t{\leq}T)$ for $x_t$:
\begin{equation}
\begin{aligned}
    &\mathbf{z}_{u,t}(2j)=\sin(t/10000^{2j/d_\text{forward}}),\\ 
    &\mathbf{z}_{u,t}(2j+1)=\cos(t/10000^{2j/d_\text{forward}}),
\end{aligned}
\end{equation}
where $j$ represents the $j$-th dimension, $d_\text{forward}$ denotes the dimension of the forward diffusion step embedding. Noted that for user $u$, the forward diffusion step embedding during the training phase is $\mathbf{z}_{u,t} + \mathbf{z}_{u,in}$, while during the inference phase it is $\mathbf{z}_{u,in}$, as the behavioral data contains inherent noise.

\subsection{FC Layer}
To map discrete user behavioral data to a continuous feature space, we employ Fully Connected (FC) layers to obtain hidden behavioral representations $\textbf{h}_u$ for each user $u$. We also incorporate a FC layer to remap the user behavioral representations back to discrete predicted behavioral data $\hat{\bm{x}}_0$. We recommend the item corresponding to the $K$ columns with the highest predicted probabilities in $\hat{\bm{x}}_0$.

\subsection{Collaboration-aware Generation Block (CG Block)}
Collaborative signals reflect similar behavioral patterns among users. To effectively leverage these common patterns alongside individual user behaviors, we propose the CG Block with a gating mechanism.

Specifically, the individual behavioral representation is first concatenated with the forward diffusion step embeddings and mapped to a hidden space:
\begin{equation}
\begin{aligned}
    \mathbf{h}_u' = \sigma(\mathbf{W}_1\mathbf[\textbf{h}_u;\mathbf{z}_{u,t}] + \mathbf{b}_1),
\end{aligned}
\end{equation}
where {$\mathbf{W}_1\in{\mathbb{R}^{d \times{(d+d_\text{forward})}}}$} and {$\mathbf{b}_1\in{\mathbb{R}^{d}}$} are learnable transformation matrix and bias matrix. $d$ denotes the dimension of the hidden layer, and {$\sigma$} corresponds to the activation function.

Next, the collaborative signals are also mapped to the same hidden space as supplementary information:
\begin{equation}
\begin{aligned}
    \mathbf{c}_u' = \sigma(\mathbf{W}_2\mathbf{c}_u + \mathbf{b}_2).
\end{aligned}
\end{equation}
where {$\mathbf{W}_2\in{\mathbb{R}^{d \times{d_{\text{svd}}}}}$} and {$\mathbf{b}_2\in{\mathbb{R}^{d}}$} are learnable transformation matrix and bias matrix. 

Then, under the control of forward diffusion step, individual behavioral and collaborative information are adaptively fused:
\begin{equation}
\begin{aligned}
    &\mathbf{h}_{u,d} = \mathbf{h}_u' \odot (1-g(\mathbf{z}_{u,t})) + \mathbf{c}_u' \odot g(\mathbf{z}_{u,t})\\
    &\;\;\;\;\;\;\;\;\;\;g(\mathbf{z}_{u,t}) = \sigma(\mathbf{W}_3\mathbf{z}_{u,t} + \mathbf{b}_3)
\end{aligned}
\end{equation}
where $g(\cdot)$ is gating function, {$\mathbf{W}_3\in{\mathbb{R}^{d \times{d}}}$} and {$\mathbf{b}_3\in{\mathbb{R}^{d}}$} are learnable parameters, and {$\odot$} represents element-wise product.

\subsection{Preference-aware Generation Block (PG Block)}
Given the association between user behaviors and personalized modality preference signals, we design the PG Block. This component  generates behaviors controlled by textual preference $\textbf{m}_{u,t}$ and visual preference $\textbf{m}_{u,v}$.

Taking generating behaviors controlled by textual modality preference signals $\textbf{m}_{u,t}$ as an example, we first utilize two MLPs to predict the parameters $\mathbf{W}_t$ for text-conditioned feature transformation and the shifting parameters:
\begin{gather}
    \mathbf{W}_g = \text{MLP}_1(\mathbf{m}_{u,t}),\quad\mathbf{b}_g = \text{MLP}_2(\mathbf{m}_{u,t}).
\end{gather}

Next, we perform a feature transformation operation on the input features $\mathbf{b}_{u,d}$ based on $\mathbf{W}_t$, and apply the channel-wise shifting operation with $\mathbf{b}_t$. This process is represented as:
\begin{gather}
    \mathbf{h}_{u,g} =  \mathbf{W}_g\mathbf{h}_{u,d} + \mathbf{b}_g.
\end{gather}

The process of behavior generation under visual preference control follows a similar way. Considering the above generation process are all linear transformation operation, we also introduce a LeakyReLU layer between textual preference-conditioned generation and visual preference-conditioned generation. This inclusion introduces nonlinearity into the generation process, expanding the representation space and enhancing the separability of representations.

\section{Experiments}
\subsection{Datasets}

\begin{table*}[h]
    \centering
    \caption{Statistics of the experimental datasets}
    \label{table:Statistics}
    \begin{tabular}{ccccc}
    \toprule
    Dataset & \#User & \#Item & \#Behavior & Density \\
    \midrule
    TMALL & 13,114 & 7,840 & 153,352 & 0.149\% \\
    Microlens & 46,420 & 14,193 & 330,683 & 0.050\% \\
    H\&m & 43,529 & 16,908 & 368,381 & 0.050\% \\
    \bottomrule
\end{tabular}
\end{table*}

We conduct experiments on three real-world datasets: (a) {TMALL}\footnote{\url{https://tianchi.aliyun.com/dataset/140281}}; (b) {Microlens}\footnote{\url{https://recsys.westlake.edu.cn/MicroLens-50k-Dataset/}};  and (c) {H\&M}\footnote{\url{https://www.kaggle.com/c/h-and-m-personalized-fashion-recommendations}}. We perform 10-core filtering on the raw data. The statistics of these datasets are presented in Table.\ref{table:Statistics}.

\subsection{Compared Methods}
To evaluate the effectiveness of our proposed model, we compare it with several representative recommendation models. These baselines fall into two groups: General recommendation methods, which only rely on interactive data for recommendation; Multimedia recommendation methods, which utilize both interactive data and multi-modal features for the recommendation.

\vspace{3pt}
\noindent \textbf{i) General Recommendation Methods:}
We have selected some of the most representative models as our compared methods, including MF-based methods (MF-BPR \cite{koren2009matrix}), GCN-based methods (LightGCN \cite{he2020lightgcn}), Diffusion-based methods (DiffRec \cite{wang2023diffusion}) and Self-Supervised Learning (SSL)-based methods (SimGCL \cite{yu2022graph}).

\vspace{3pt}
\noindent \textbf{ii) Multimedia Recommendation Methods:}
To enhance the evaluation of our approach, we selected some competitive models as our compared methods. This encompasses techniques such as MF-based methods (VBPR \cite{he2016vbpr}), GCN-based methods (LATTICE \cite{zhang2021mining}, MGCN \cite{yu2023multi}), SSL-based methods (SLMRec \cite{tao2022self}, BM3 \cite{zhou2023bootstrap}, MMSSL \cite{wei2023multi}) and Diffusion-based methods (DiffMM \cite{jiang2024diffmm}).

\subsection{Evaluation Protocols}
To ensure a a fair comparison, we adhere to the standardized all-ranking protocol \cite{yu2023multi}, and select Recall@{$K$} (R@{$K$}) and NDCG@{$K$} (N@{$K$}) as the evaluation indicators. We repeat the experiment more than ten times and report average results. The significance testing results indicate that our proposed approach outperforms the best baseline significantly ($p$-value $<$ 0.05).

\subsection{Implementation Details}
The proposed framework and all compared methods are implemented using the MMRec framework\footnote{\url{https://github.com/enoche/MMRec}} \cite{zhou2023mmrecsm}, which is a unified open-source platform for developing and reproducing recommendation algorithms. To ensure a fair comparison, all methods are optimized using the Adam optimizer, and hyperparameters follow the settings reported in their original papers.

For DiffRec and our method, we set the input batch size to 400. To reduce the search space for hyperparameters in LD4MRec, we fix the embedding size $d$ at 1000, the $d_\text{svd}$ size at 100, and the $d_\text{forward}$ size at 10. For the noise introduced in the forward diffusion process, we set the noise scale $s=0.1$ and the minimum level of noise $\alpha_{min}=0.0001$. And we perform a grid search on the remaining hyperparameters of LD4MRec across all datasets to determine their optimal values. Specifically, for the soft reconstruction loss, we search for the optimal value of $p$ from the set \{0.005, 0.01, 0.05, 0.1, 0.2\}, and the optimal value of $\gamma$ from the set \{0.005, 0.01, 0.05, 0.1, 0.2\}. To ensure convergence, we have employed early stopping after 20 epochs and a total of 1000 epochs. Following the approach described in \cite{zhang2022latent}, we have utilized Recall@20 on the validation data as the indicator for stopping the training process. All experiments are performed using PyTorch on NVIDIA Tesla V100 GPUs.

\subsection{Performance Comparison}
\begin{table*}[t]
\centering
\caption{Performance Comparison on TMALL Dataset.}
\label{tab:tmall_results}
\setlength{\tabcolsep}{4.0mm}{
\begin{tabular}{c|cccc}
\toprule
\textbf{Methods} & \textbf{R@10} & \textbf{R@20}  & \textbf{N@10}  & \textbf{N@20}  \\ \midrule
MF-BPR & 0.0157  & 0.0232  & 0.0100  & 0.0121  \\
LightGCN & 0.0228  & 0.0345  & 0.0133  & 0.0165  \\
DiffRec & 0.0229  & 0.0340  & 0.0136  & 0.0159  \\
SimGCL & 0.0233  & 0.0355  & 0.0135  & 0.0169  \\ \midrule
VBPR & 0.0221  & 0.0345  & 0.0121  & 0.0159  \\
LATTICE & 0.0238  & 0.0356  & 0.0130  & 0.0167  \\
SLMRec & 0.0237  & 0.0354  & 0.0128  & 0.0166  \\
BM3 & 0.0234  & 0.0351  & 0.0128  & 0.0163  \\ 
MMSSL & 0.0243  & 0.0367  & 0.0133  & 0.0169  \\
MGCN & \underline{0.0249}  & \underline{0.0380}  & \underline{0.0135} & \underline{0.0171} \\ 
DiffMM & 0.0241 & 0.0364 & 0.0131 & 0.0169 \\ \midrule
\textbf{LD4MRec} & \textbf{0.0263}  & \textbf{0.0409}  & \textbf{0.0147}  & \textbf{0.0190}  \\
\textbf{$p$-value} & 5.89e-5 & 6.01e-5 & 2.11e-4 & 2.32e-4 \\
\bottomrule
\end{tabular}
}
\end{table*}

\begin{table*}[t]
\centering
\caption{Performance Comparison on Microlens Dataset.}
\label{tab:microlens_results}
\setlength{\tabcolsep}{4.0mm}{
\begin{tabular}{c|cccc}
\toprule
\textbf{Methods} & \textbf{R@10} & \textbf{R@20}  & \textbf{N@10}  & \textbf{N@20}  \\ \midrule
MF-BPR & 0.0479  & 0.0794  & 0.0258  & 0.0347  \\
LightGCN & 0.0498  & 0.0820  & 0.0273  & 0.0364  \\
DiffRec & 0.0511  & 0.0847  & 0.0277  & 0.0373  \\
SimGCL & 0.0523  & 0.0861  & 0.0291  & 0.0388  \\ \midrule
VBPR & 0.0515  & 0.0854  & 0.0288  & 0.0384  \\
LATTICE & 0.0553  & 0.0886  & 0.0308  & 0.0402  \\
SLMRec & 0.0569  & 0.0922  & 0.0314  & 0.0415  \\
BM3 & 0.0537  & 0.0877  & 0.0293  & 0.0398  \\ 
MMSSL & 0.0576 & 0.0936 & 0.0320 & 0.0422 \\
MGCN & 0.0584  & 0.0941 & 0.0327 & 0.0426 \\ 
DiffMM & \underline{0.0590} & \underline{0.0954} & \underline{0.0328} & \underline{0.0432} \\ \midrule
\textbf{LD4MRec} & \textbf{0.0618} & \textbf{0.0972} & \textbf{0.0342} & \textbf{0.0442} \\
\textbf{$p$-value} & 7.89e-5 & 7.96e-5 & 3.72e-4 & 3.81e-4 \\
\bottomrule
\end{tabular}
}
\end{table*}

\begin{table*}[t]
\centering
\caption{Performance Comparison on H\&M Dataset.}
\label{tab:hm_results}
\setlength{\tabcolsep}{4.0mm}{
\begin{tabular}{c|cccc}
\toprule
\textbf{Methods} & \textbf{R@10} & \textbf{R@20}  & \textbf{N@10}  & \textbf{N@20}  \\ \midrule
MF-BPR & 0.0209 & 0.0320  & 0.0121 & 0.0150 \\
LightGCN & 0.0235 & 0.0360  & 0.0139 & 0.0172 \\
DiffRec & 0.0244  & 0.0368  & 0.0142  & 0.0175  \\
SimGCL & 0.0252  & 0.0375  & 0.0146  & 0.0178  \\ \midrule
VBPR & 0.0241 & 0.0370  & 0.0138 & 0.0175  \\
LATTICE & 0.0289 & 0.0427  & 0.0161 & 0.0187  \\
SLMRec & 0.0308 & 0.0464  & 0.0179 & 0.0220  \\
BM3 & 0.0294 & 0.0434 & 0.0165 & 0.0192 \\ 
MMSSL & 0.0327 & 0.0503  & 0.0185  & 0.0229  \\
MGCN & 0.0340 & 0.0522 & 0.0190 & 0.0238  \\ 
DiffMM & \underline{0.0367}  & \underline{0.0549}  & \underline{0.0204} & \underline{0.0251} \\ \midrule
\textbf{LD4MRec} & \textbf{0.0381} & \textbf{0.0573} & \textbf{0.0211} & \textbf{0.0269} \\
\textbf{$p$-value} & 1.28e-5 & 1.39e-5 & 2.67e-4 & 2.78e-4 \\
\bottomrule
\end{tabular}
}
\end{table*}

\noindent $\bullet$ \textbf{Effectiveness:} According to Table \ref{tab:tmall_results}, Table \ref{tab:microlens_results} and Table \ref{tab:hm_results}, LD4MRec outperforms all the compared methods across three real-world datasets. On the TMALL dataset, it improves Recall@20 by 7.6\% and NDCG@20 by 11.1\%. The performance gap becomes more pronounced on the sparser Microlens dataset, where LD4MRec outperforms the strongest baseline DiffMM by 4.7\% in Recall@20 and 2.3\% in NDCG@20. These improvements validate our core hypothesis that conditional generation guided by collaborative and modality preference signals can better align generated behaviors with user preferences. We attribute this superiority to three key factors: (1) Unlike GNN-based methods like MGCN that amplify noise through message passing, or SSL-based approaches like MMSSL that risk semantic distortion through random perturbation, LD4MRec explicitly models behavioral uncertainty. The soft reconstruction loss further enables robust preference extraction from noisy observations. (2) The dual conditional generation  provides complementary guidance. Collaborative signals capture macroscopic behavioral patterns (e.g., user clusters preferring specific product categories), while modality preferences model microscopic decision-making mechanisms (e.g., visual-driven purchases). (3) The gating mechanism in CG Block automatically adjusts the fusion ratio between collaborative and individual signals based on diffusion steps. This allows preserving high-confidence historical interactions while supplementing potential interests during generation. Thereby, LD4MRec demonstrates improved recommendation performance.

\noindent $\bullet$ \textbf{Efficiency:} We present the training and inference time in Table \ref{tab:efficiency}. Our results show that LD4MRec largely reduces both the training and inference time than existing diffusion-based recommendation models. On the largest H\&M dataset,  LD4MRec achieves a 93.2\% reduction in inference time relative to DiffRec, while maintaining comparable training efficiency. This improvement is attributed to two key factors: (1) The elimination of the iterative corruption process during inference, which removes the $O(T)$ computational complexity, where T=1000 in typical diffusion settings. (2) The stacked CG/PG blocks employ shallow transformations with dimension compression, resulting in fewer FLOPs compared to DiffMM's cross-modal contrastive modules.

\begin{table}[t]
\centering
\caption{Efficiency Comparison of Different Recommendation Models in terms of Time (s/epoch).}
\label{tab:efficiency}
\begin{tabular}{ccccccc}
\toprule
\textbf{Dataset} & \textbf{Metric} & \textbf{MMSSL} & \textbf{MGCN} & \textbf{DiffRec} & \textbf{DiffMM} & \textbf{LD4MRec} \\ \midrule
 \multirow{3}{*}{\textbf{TMALL}} & \textbf{Training} & 2.73 & 2.19 & 7.88 & 9.19 & 8.61    \\
  & \textbf{Inference} & 2.01 & 1.72 & 112.57 & 30.62 & 8.99  \\ 
  & \textbf{Memory} & 3.17 & 1.99 & 1.41 & 2.84 & 2.58 \\
  \midrule
  \multirow{3}{*}{\textbf{Micreolens}} & \textbf{Training} & 4.28 & 3.85 & 12.72 & 15.11 & 10.20    \\
  & \textbf{Inference} & 2.68 & 2.17 & 251.42 & 173.61 & 14.37  \\ 
  & \textbf{Memory} & 4.65 & 3.62 & 2.37 & 4.20 & 3.91 \\
  \midrule
  \multirow{2}{*}{\textbf{H\&M}} & \textbf{Training} & 4.89 & 4.33 & 14.06 & 16.21 & 12.94    \\
  & \textbf{Inference} & 3.70 & 3.44 & 294.81 & 199.15 & 20.08  \\ 
  & \textbf{Memory} & 4.39 & 3.46 & 2.28 & 4.11 & 3.67 \\
\bottomrule \\
\end{tabular}
\end{table}

\subsection{In-depth Analysis}
\begin{figure}[t]
  \centering
  \includegraphics[width=\linewidth]{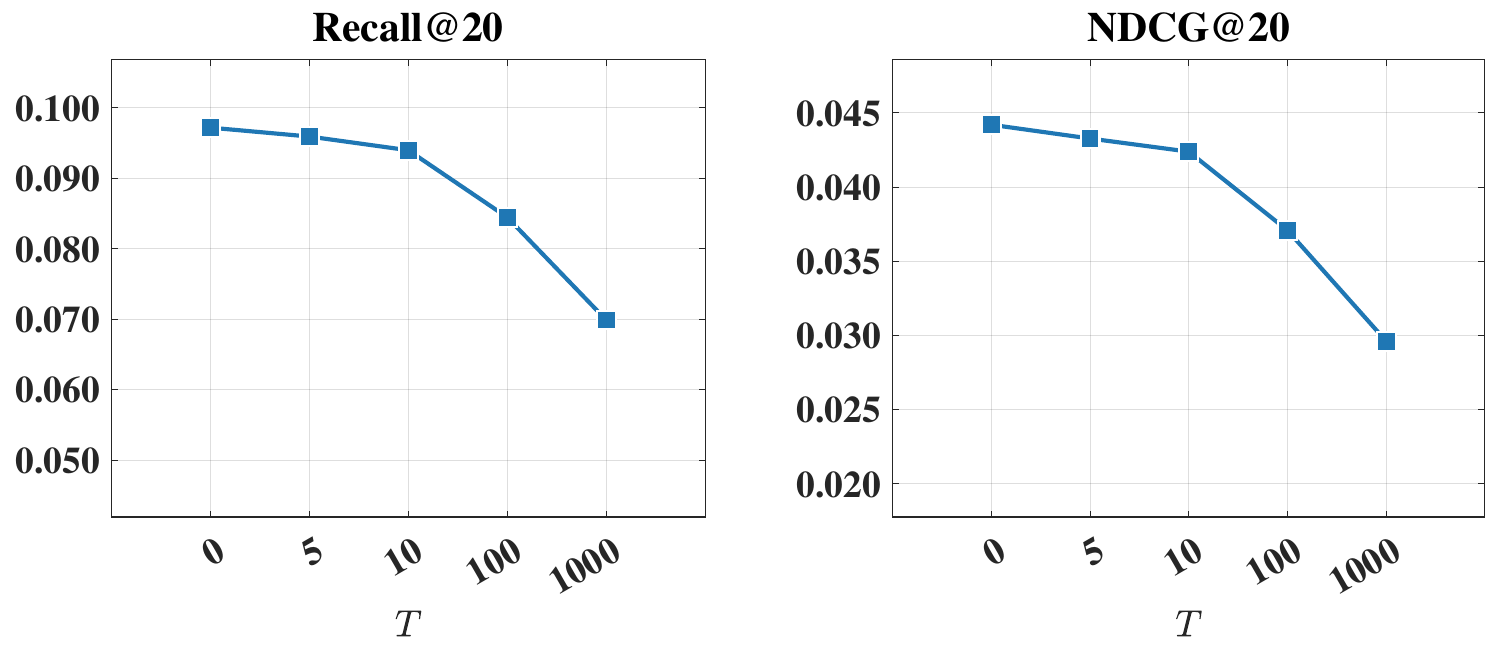}
  \caption{Performance comparison between different forward steps.}
  \label{fig.forward}
\end{figure}

\noindent $\bullet$ \textbf{Effect of Forward-free Inference:} We compare the proposed forward-free inference strategy with different forward step inference strategies. The results, presented in Fig. \ref{fig.forward}, indicate that the forward-free inference strategy outperforms the others. This is because forward-free inference strategy preserves valuable behavioral semantic information, and avoids excessive corrupting of already sparse historical behaviors.

\begin{figure}[t]
  \centering
  \includegraphics[width=\linewidth]{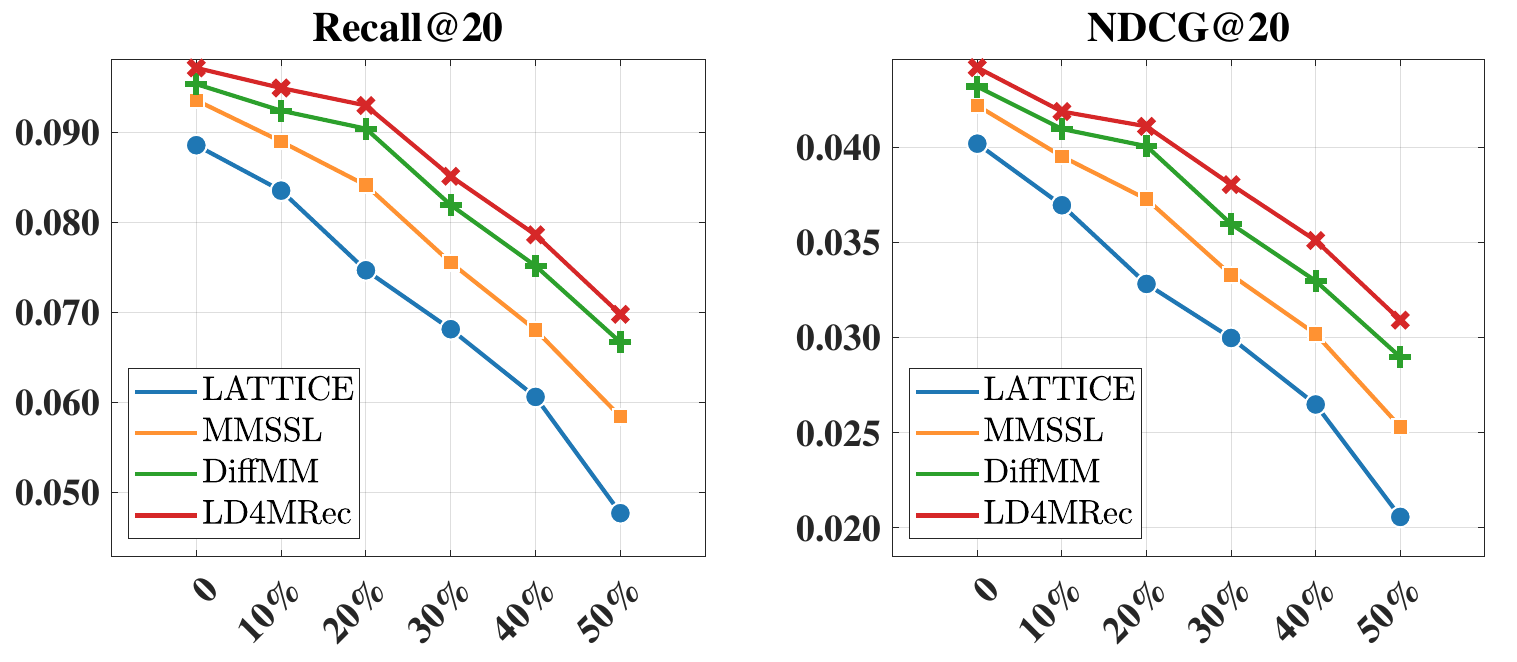}
  \caption{Performance comparison between different noise levels.}
  \label{fig.robust}
\end{figure}

\begin{figure}[t]
  \centering
  \includegraphics[width=\linewidth]{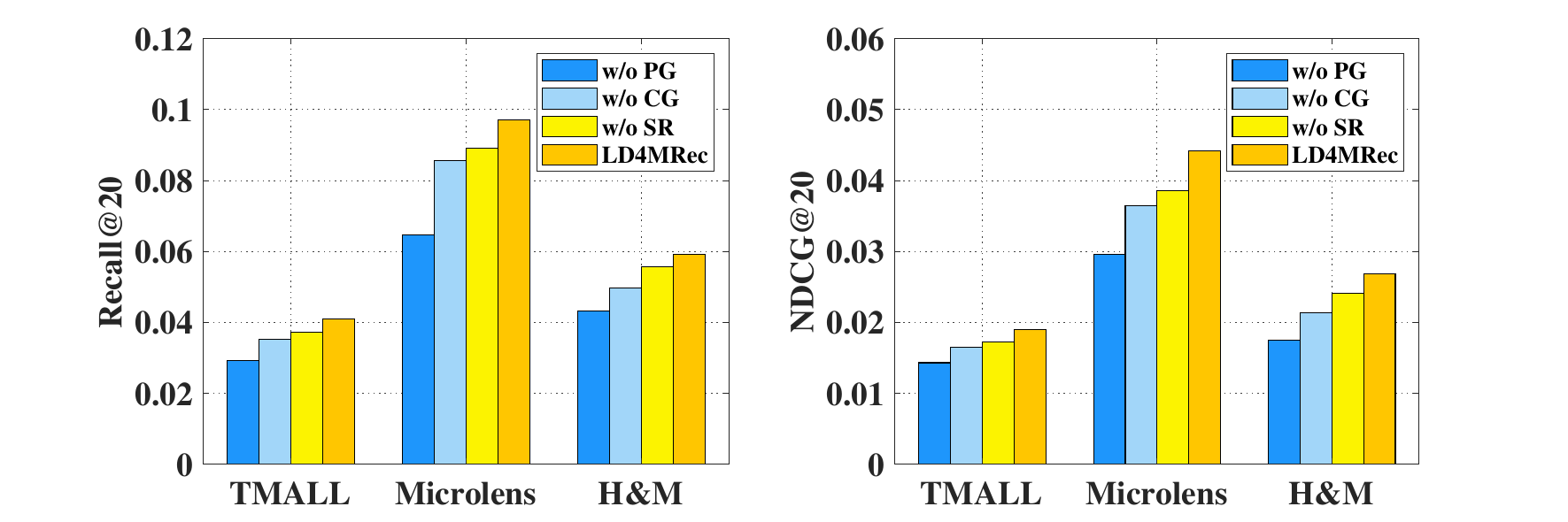}
  \caption{Performance comparison between different variants.}
  \label{fig.ablation}
\end{figure}

\vspace{3pt}
\noindent $\bullet$ \textbf{Robustness of LD4MRec:} To assess the robustness of LD4MRec, we introduce varying levels of noise into the observed user behaviors. As shown in Fig. \ref{fig.robust}, LD4MRec exhibited the best performance under different proportions of noise. Diffusion-based multimedia recommendation methods (DiffMM, LD4MRec) demonstrate relatively little performance degradation. This is primarily because the inherent ability of diffusion models to resist noise. In contrast, GNN-based recommendation method (LATTICE) experiences the most significant performance decline, as the message passing mechanism in graph neural networks amplifies the harm of behavioral noise. While SSL-based method (MMSSL) alleviates the impact of noise by designing self-supervised learning auxiliary tasks, the noise still affects the performance.

\vspace{3pt}
\noindent $\bullet$ \textbf{Effects of key module:}
We further evaluate the contributions of each key module and present the results in Fig. \ref{fig.ablation}. "w/o PG" indicates that we exclude the PG blocks and sequentially link the two CG blocks. "w/o CG" indicates that CG block are eliminated. Similar to DiffRec \cite{wang2023diffusion}, behavioral features is concatenated with forward diffusion step embedding, and then fed to a MLP. "w/o SR" denotes the omission of soft reconstruction loss and employs the classic MSE loss. Our findings indicate that the removal of any module results in a decline in performance, particularly when the CG and PG blocks are excluded. This underscores the importance of dual guidance signals in achieving the conditional generation of behavioral information. Additionally, the soft reconstruction loss also plays a critical role. It allows some flexibility and encourages the extraction of invariant user preferences.

\vspace{3pt}
\noindent $\bullet$ \textbf{Hyperparameter Analysis:} 
\begin{figure}
  \centering
  \includegraphics[width=\linewidth]{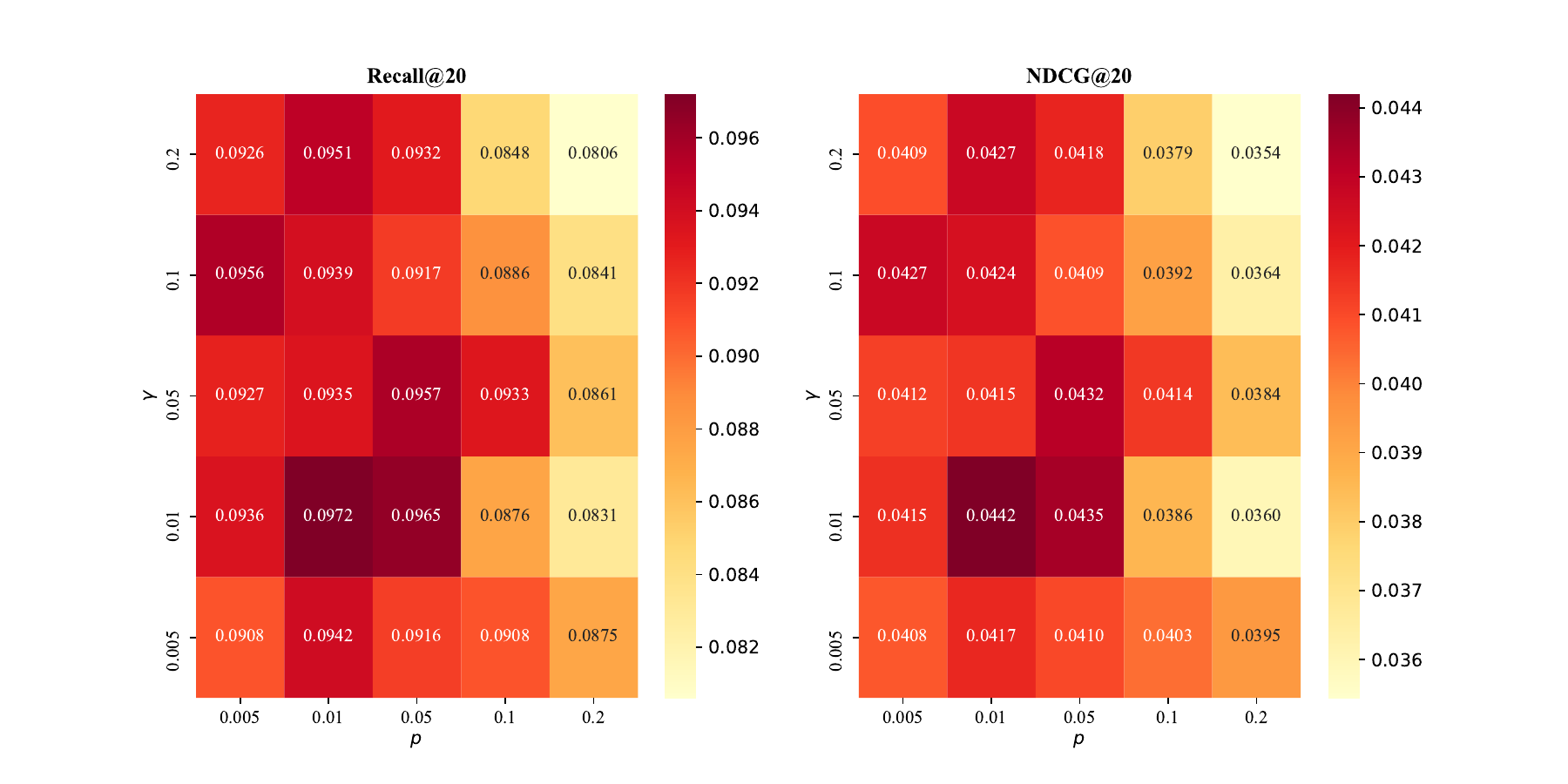}
  \caption{Performance comparison w.r.t. different $p$ and $\gamma$.}
  \label{fig.hyper_pv}
\end{figure}
To facilitate the effective learning of the model from noise, we devise a soft reconstruction loss. We determine that optimal outcomes are obtained when the soft probability $p$ is set to 0.01, and the smoothing intensity $\gamma$ is set to 0.01. Excessive values for either the soft probability or smoothing intensity can result in the loss of behavior information, leading to suboptimal results.

\section{Conclusion}
In this paper, we have proposed a Light Diffusion model for Multimedia Recommendation ({\bfseries LD4MRec}). Specifically, we greatly simplify the conventional Diffusion Models for multimedia recommendation. And we design a C-Net to fit the reverse process and effectively generate personalized behavioral information.

In our future work, we will focus on optimizing the design of the C-Net to enhance the quality of generation. Additionally, we plan to explore the diffusion of behavioral information in the hidden space, with the intention of reducing the number of parameters and computational requirements.

\nocite{*}
\bibliography{sn-bibliography}

\end{document}